\documentstyle[12pt,dmtrieste,psfig]{article}

\def\mincir{\raise -2.truept\hbox{\rlap{\hbox{$\sim$}}\raise5.truept
\hbox{$<$}\ }}
\def\magcir{\raise -2.truept\hbox{\rlap{\hbox{$\sim$}}\raise5.truept
\hbox{$>$}\ }}

\catcode`\@=11
\long\def\@makefntext#1{
\protect\noindent \hbox to 3.2pt {\hskip-.9pt
$^{{\ninerm\@thefnmark}}$\hfil}#1\hfill}                

\def\@makefnmark{\hbox to 0pt{$^{\@thefnmark}$\hss}}  

\def\ps@myheadings{\let\@mkboth\@gobbletwo
\def\@oddhead{\hbox{}
\rightmark\hfil\ninerm\thepage}
\def\@oddfoot{}\def\@evenhead{\ninerm\thepage\hfil
\leftmark\hbox{}}\def\@evenfoot{}
\def\sectionmark##1{}\def\subsectionmark##1{}}

\newcommand{\lsim}{\mathrel{\mathop{\kern 0pt \rlap
  {\raise.2ex\hbox{$<$}}}
  \lower.9ex\hbox{\kern-.190em $\sim$}}}
\newcommand{\gsim}{\mathrel{\mathop{\kern 0pt \rlap
  {\raise.2ex\hbox{$>$}}}
  \lower.9ex\hbox{\kern-.190em $\sim$}}}

\begin{document}

\thispagestyle{empty}

\rightline{DFTT 18/98}
\rightline{January 1998}

\vspace{3pc}
\centerline{\Large \bf SUPERSYMMETRIC CANDIDATES}
\centerline{\Large \bf  FOR DARK MATTER
\footnote{Report on the work done in collaboration with A. Bottino, F. Donato and S. Scopel.}}

\vspace{3pc}
\centerline{\large \bf Nicolao Fornengo}

\vspace{1pc}
\em
\begin{center}
\begin{tabular}{c}
Dipartimento di Fisica Teorica, Universit\`a di Torino \\
and \\
INFN, Sezione di Torino \\
Via P. Giuria 1, 10125 Torino, Italy
\\
{\sl fornengo@to.infn.it}
\\
{\sl http://www.to.infn.it/$\tilde{\hspace{1ex}}$fornengo/index.html}

\end{tabular}
\end{center}

\vspace{3pc}
\rm
\centerline{\large \bf Abstract}
\bigskip

Direct and indirect detection rates of relic neutralinos are 
reviewed in the framework of the Minimal Supersymmetric
Standard Model. The theoretical estimates are compared with the most 
recent experimental limits from low--background detectors and
neutrino telescopes. The properties of neutralino under the hypothesis that 
preliminary experimental results of the DAMA/NaI Collaboration  
may be indicative of a yearly modulation effect are examined.
\vfill

\begin{center}
{ \em
  Talk presented at 
  ``DM97: Dark matter: perspectives and projects'', 
   Osservatorio Astronomico and ICTP, Trieste, 9--11 December 1997}
\end{center}
\begin{center}
(published in the Proceedings, editor P. Salucci)
\end{center}

\vspace{2pc}
\eject
\setcounter{page}{1}

\centerline{\normalsize\bf SUPERSYMMETRIC CANDIDATES FOR DARK MATTER
\footnote{Report on the work 
done in collaboration with A. Bottino, F. Donato and S. Scopel.}}

\vspace*{0.6cm}
\centerline{\footnotesize NICOLAO FORNENGO}
\baselineskip=13pt
\centerline{\footnotesize\it Dipartimento di Fisica Teorica, Universit\`a di Torino}
\baselineskip=12pt
\centerline{\footnotesize \it and INFN - Sezione di Torino}
\centerline{\footnotesize\it via P. Giuria 1, 10125 Torino, Italy}
\centerline{\footnotesize fornengo@to.infn.it}

\vspace*{0.9cm}
\abstracts{Direct and indirect detection rates of relic neutralinos are 
reviewed in the framework of the Minimal Supersymmetric
Standard Model. The theoretical estimates are compared with the most 
recent experimental limits from low--background detectors and
neutrino telescopes. The properties of neutralino under the hypothesis that 
preliminary experimental results of the DAMA/NaI Collaboration  
may be indicative of a yearly modulation effect are examined.}

\normalsize\baselineskip=15pt

\vspace{-5pt}
\section{Introduction}
\vspace{-5pt}

Supersymmetric theories predict a large number of particles in
excess to the Standard Model ones. If the R--parity is conserved,
the lightest among all the supersymmetric particles (LSP) must be stable. This
feature makes the LSP a dark matter candidate, since this particle can
be present today as a relic from the early stages of the evolution of the 
Universe. Different candidates have been proposed in the framework of 
supersymmetric theories: the neutralino or the sneutrino\cite{sneutrino} in gravity 
mediated models, the gravitino\cite{gravitino} or some messenger fields 
in gauge mediated theories\cite{messenger},
the axino\cite{axino}, stable non--topological solitons (Q--balls)\cite{Qballs} or others.

In this paper we will focus on the most promising among all
the different candidates, the neutralino, which
is defined as the lowest mass
linear superposition of photino ($\tilde \gamma$),
zino ($\tilde Z$) and the two higgsino fields
($\tilde H_1^{\circ}$, $\tilde H_2^{\circ}$), i.e.
$\chi \equiv a_1 \tilde \gamma + a_2 \tilde Z + a_3 \tilde H_1^{\circ}  
+ a_4 \tilde H_2^{\circ}$.
The aim of this review is at providing the latest results on the
calculation of different kinds of detection rates of relic neutralinos,
in the framework of the Minimal Supersymmetric extension of the Standard Model (MSSM),
constrained by the most recent experimental data coming from accelerator physics.
We do not discuss here the details of the model, for which we refer to
Refs.\cite{pinning,extending} and to the references quoted therein.
We only recall the standard assumptions employed here:
i) all trilinear parameters are set to zero except those of the third family, 
which are unified to a common value $A$;
ii) all squarks and sleptons soft--mass parameters are taken as 
degenerate: $m_{\tilde l_i} = m_{\tilde q_i} \equiv m_0$;
iii) the gaugino masses are assumed to unify at $M_{GUT}$, and this implies
$M_1= (5/3) \tan^2 \theta_W M_2$ at the electroweak scale.
After these conditions are applied, the free parameters are:
$M_2, \mu, \tan\beta, m_A, m_0, A$. The parameters are varied in
the following ranges: $10\;\mbox{GeV} \leq M_2 \leq  500\;\mbox{GeV},\; 
10\;\mbox{GeV} \leq |\mu| \leq  500\;\mbox{GeV},\;
65\;\mbox{GeV} \leq m_A \leq  500\;\mbox{GeV},\; 
100\;\mbox{GeV} \leq m_0 \leq  500\;\mbox{GeV},\;
-3 \leq {\rm A} \leq +3,\;
1.01 \leq \tan \beta \leq 50$. 
In our analysis the supersymmetric parameter space is constrained by
all the experimental limits on Higgs, neutralino, chargino and sfermion
searches at accelerators. Moreover, the constraints 
due to the $b \rightarrow s + \gamma$ process\cite{LEP} are satisfied.
In addition to the experimental limits, we require
that the neutralino is the lightest 
supersymmetric particle. Finally, the regions of the parameter space where
the neutralino relic abundance exceeds the cosmological bound, i.e. 
$\Omega_{\chi}h^2 > 1$, are also excluded.

\vspace{-7pt}
\section{Direct detection} 
\label{sec:direct}
\vspace{-10pt}
\begin{figure}[t]
\centerline{\psfig{figure=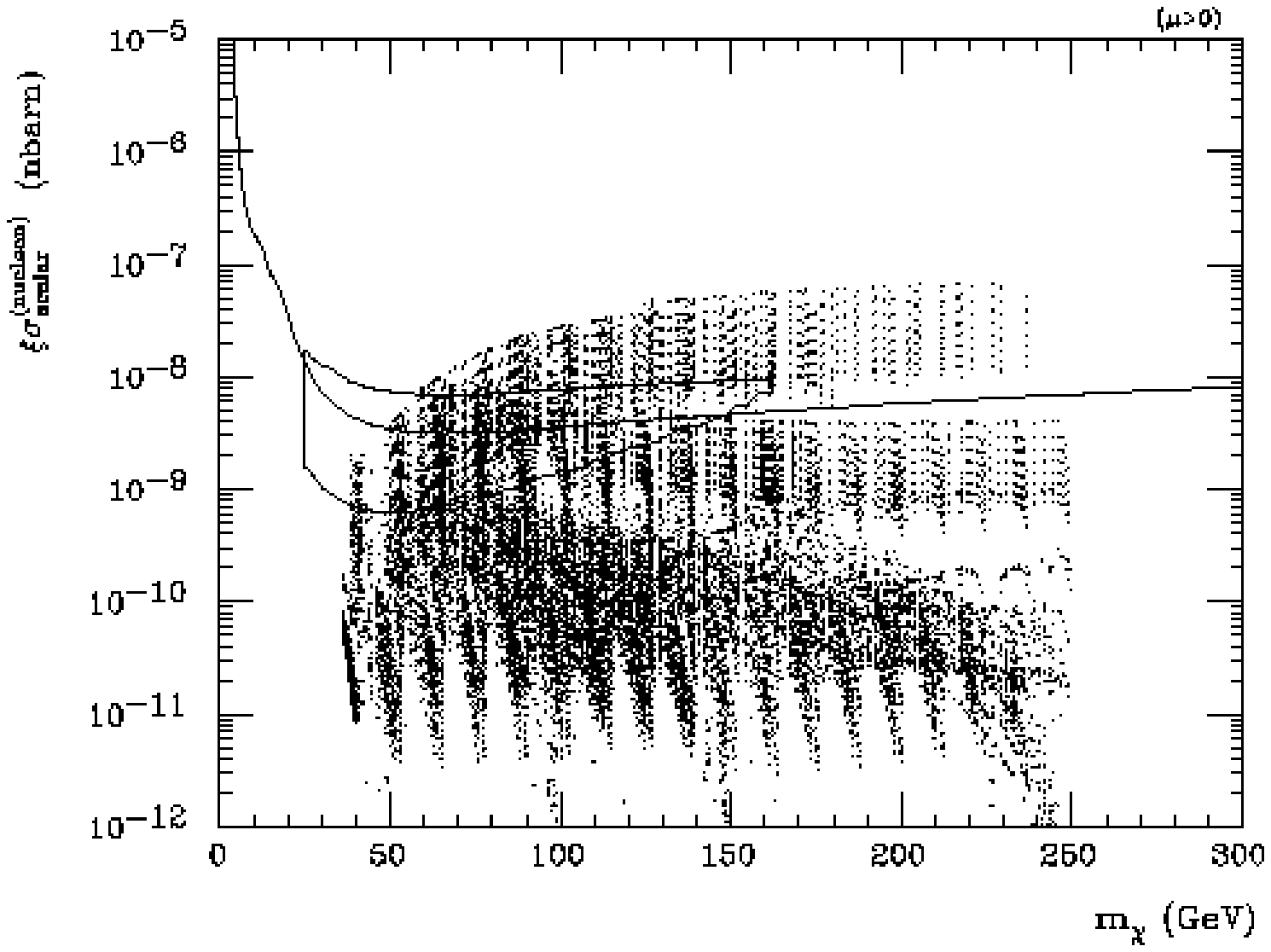,width=4.65in,bbllx=36bp,bblly=225bp,bburx=576bp,bbury=576bp,clip=}}
\fcaption{The scalar neutralino--nucleon cross section 
$\sigma^{\rm (nucleon)}_{\rm scalar}$, multiplied by the 
scaling factor $\xi$, is plotted versus the neutralino mass $m_\chi$.
The open curve denotes the 90\% C.L. upper bound 
obtained from the total counting rates of Ref.$^6$.
The scatter plot represents the theoretical predictions
calculated within the MSSM scheme ($\mu>0$ only).
The closed contour delimits the region 
singled out at 90\% C.L. when the data of Ref.$^{12}$
are interpreted in terms of a modulation signal. \label{fig:sigma}}
\end{figure}

Neutralinos interact with matter both through coherent 
and spin dependent effects \cite{pinning,diretta}.
We confine our discussion to the coherent effects, since 
these are the ones which are currently accessible to direct detection \cite{diretta}.
The relevant quantity, dependent on the supersymmetric parameters, which 
enters in the event rate of direct detection as well as in the indirect 
signals considered in Sect.3, is $\rho_\chi \times 
\sigma^{(\rm nucleon)}_{\rm scalar}$, where 
$\sigma^{(\rm nucleon)}_{\rm scalar}$ is the $\chi$--nucleon scalar cross--section
and $\rho_\chi$ is the neutralino local density. 
The expression of $\sigma^{(\rm nucleon)}_{\rm scalar}$ and its relation to
the differential event rate for elastic neutralino--nucleus scattering 
may be found in Ref.\cite{diretta}. The $\chi$ local density
can be factorized as $\rho_\chi = \xi \rho_l$, i.e. in terms of the total local dark 
matter density $\rho_l$. Here $\xi$ is evaluated as 
$\xi = {\rm min}(1,\Omega_\chi h^2 / (\Omega h^2)_{\rm min})$, where
$\Omega_\chi h^2$ is calculated as a function of the
supersymmetric parameters\cite{pinning,omega} and
$(\Omega h^2)_{\rm min}$ is a minimal value compatible with observational data and with large--scale 
structure calculations\cite{extending}. All the results of this paper refer to the choice
$\rho_l = 0.5$ GeV cm$^{-3}$ and $(\Omega h^2)_{\rm min} = 0.03$\cite{pinning}.

Fig.1 shows the present most stringent 90\% C.L. upper limit on the quantity
$\xi\sigma^{\rm (nucleon)}_{\rm scalar}$, obtained by
the DAMA/NaI Collaboration\cite{damapsd}. A possible rotation of the
galactic halo can affect the exclusion plot less than a factor of two
for $m_\chi \gsim 50$ GeV\cite{rotation}. In the same figure
our results are provided in the form of a scatter plot,
obtained by varying the parameters of the supersymmetric model in
the intervals quoted above (only $\mu > 0$ is displayed).
We see that the scatter plot of the theoretical predictions reaches 
abundantly the curve of the 90\% C.L. upper bound.
This shows that the sensitivity of the direct detection experiment is adequate 
for a significant exploration of the neutralino parameter space. 
This result applies to the above mentioned choice of the
astrophysical parameters $\rho_l$ and $(\Omega h^2)_{\rm min}$.
Different choices of these parameters inside their allowed 
ranges\cite{pinning,diretta} can change the theoretical predictions
roughly by one order of magnitude.

\vspace{-7pt}
\section{Indirect detection at neutrino telescopes}
\label{sec:indirect}
\vspace{-5pt}
\begin{figure}[t]
\centerline{\psfig{figure=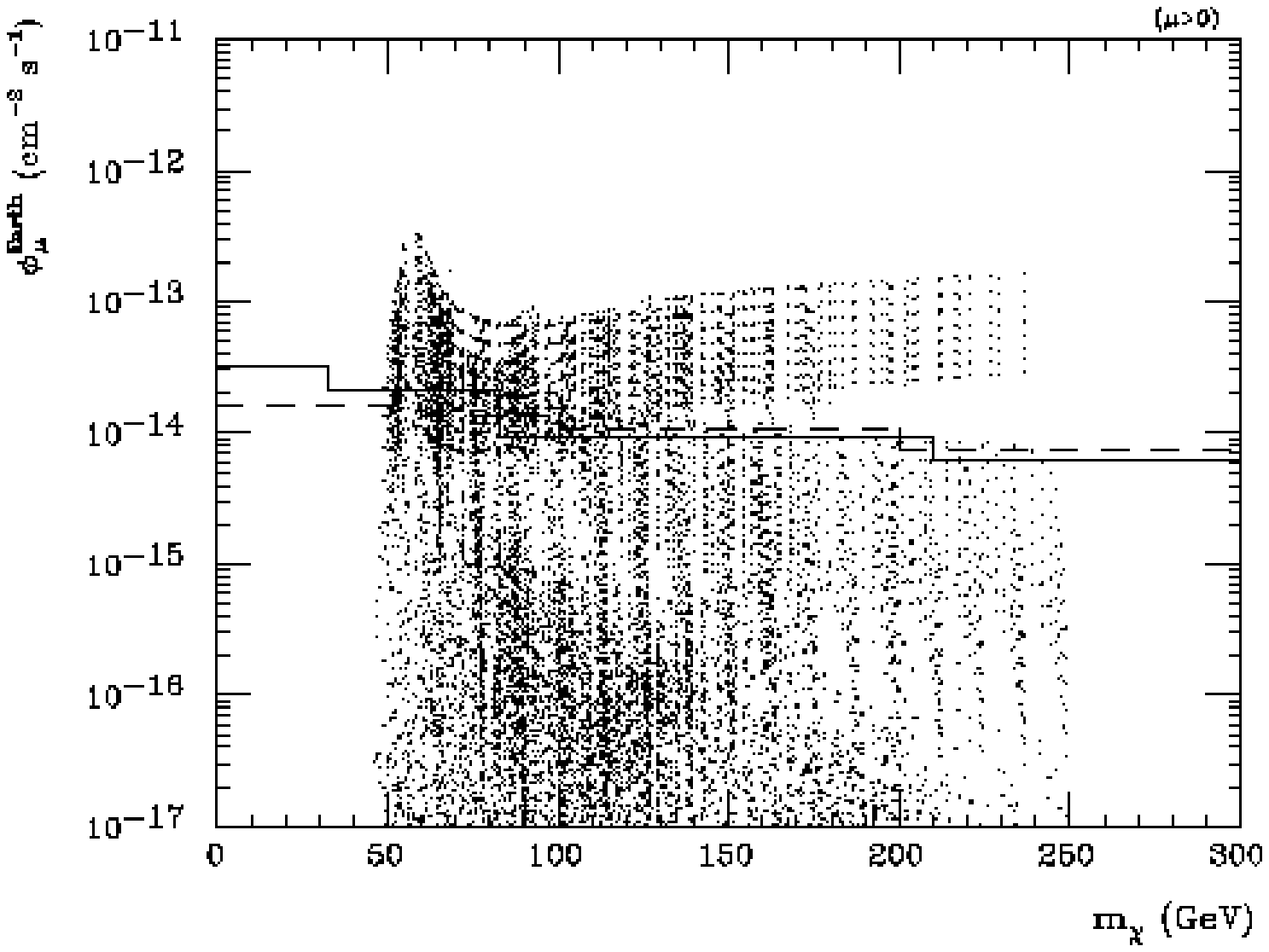,width=4.65in,bbllx=36bp,bblly=225bp,bburx=576bp,bbury=576bp,clip=}}
\fcaption{Flux of up-going muons $\Phi_{\mu}^{\rm Earth}$
as a function of $m_{\chi}$, calculated within the MSSM scheme ($\mu>0$).
The solid (dashed) line represents the experimental 90\% C.L. upper bound of 
Ref.$^9$ ($^{10}$). \label{fig:flux_E}}
\end{figure}

\begin{figure}[t]
\centerline{\psfig{figure=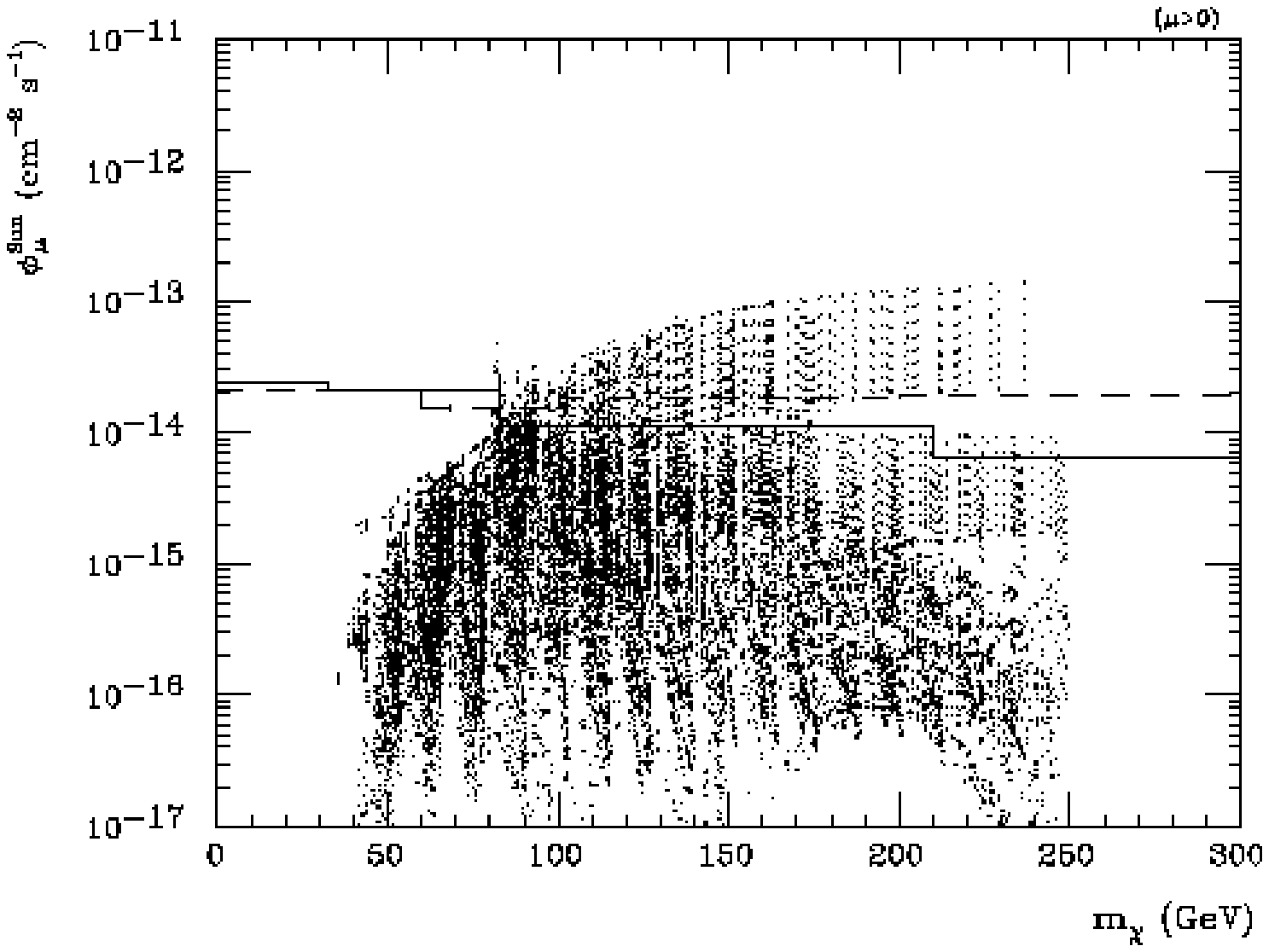,width=4.65in,bbllx=36bp,bblly=225bp,bburx=576bp,bbury=576bp,clip=}}
\fcaption{Flux of up-going muons $\Phi_{\mu}^{\rm Sun}$
as a function of $m_{\chi}$, calculated within the MSSM scheme ($\mu>0$).
The solid (dashed) line represents the experimental 90\% C.L. upper bound of 
Ref.$^9$ ($^{10}$). \label{fig:flux_S}}
\end{figure}

Pair annihilations of neutralinos may provide other signals for the relic
neutralino searches. The annihilation process may occur in the galactic halo
or inside celestial bodies, like the Earth or the Sun, where the neutralino may 
have been accumulated as a consequence of gravitational capture. In the case of 
annihilation in the halo, the signal consists of photon\cite{lars}, 
positron and antiproton\cite{pbar} fluxes,
which can be observed by detectors placed on balloons or satellites.
Recent calculations of the $\bar p$ fluxes in the MSSM will be presented elsewhere\cite{review}.
In the case of neutralinos captured in the Earth and in the Sun, the signal
is a flux of $\nu_\mu$'s which can be detected as up--going muons
in a neutrino telescope. The values of the calculated fluxes of up--going 
muons\cite{bere,indiretta} $\Phi_\mu^{\rm Earth}$ and $\Phi_\mu^{\rm Sun}$ are displayed in 
Figs.2,3 together with the experimental 90\% C.L. upper bounds of 
Ref.\cite{baksan} (solid line) and Ref.\cite{macro} (dashed line).
From Figs.2,3 it turns out that many supersymmetric configurations 
give muon fluxes exceeding the present experimental upper bounds.
Also in this case, the variation of the astrophysical parameters
may change the calculated fluxes roughly by one order of magnitude.

Another possible indirect signal, relevant to the neutrino telescope searches, 
may consist in a $\nu_\mu$ flux produced by neutrino annihilation inside
the Large Magellanic Cloud (LMC). 
Theoretical estimates of the neutrino signal from LMC will be presented
elsewhere\cite{salucci}.

\vspace{-7pt}
\section{Direct detection: yearly modulation of the signal}
\vspace{-5pt}
\begin{figure}[t]
\centerline{\psfig{figure=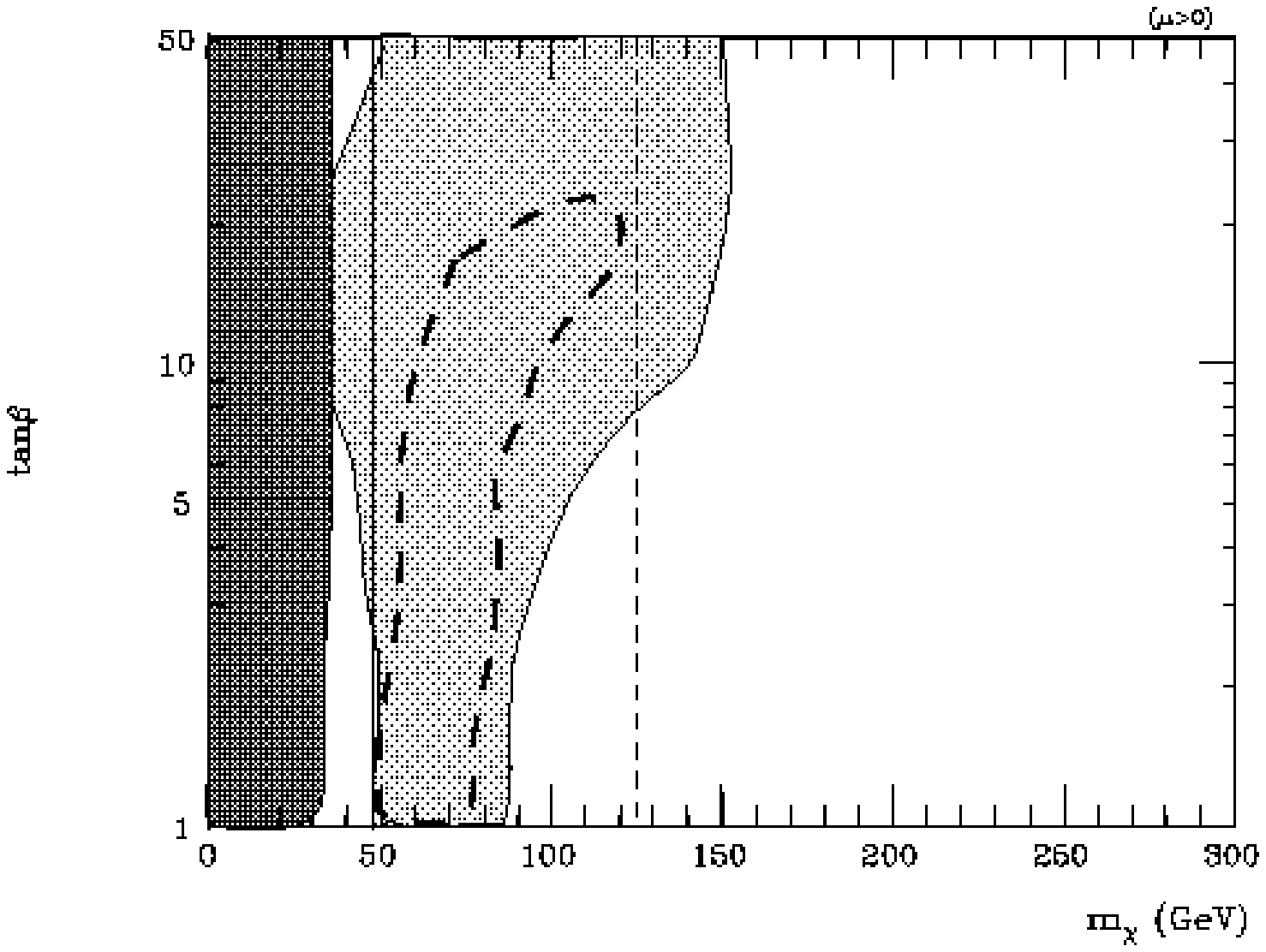,width=4.65in,bbllx=36bp,bblly=272bp,bburx=576bp,bbury=626bp,clip=}}
\fcaption{The configurations compatible with the closed region
of Fig. 1 are plotted in the $m_\chi$--$\tan\beta$ plane, within the gray area. 
Configurations which provide muon fluxes from the Earth and from
the Sun which exceed the limits of Refs.$^{9,10}$ have been dropped.
The dark region on the left side is 
excluded by current LEP data$^3$. The region on the left
of the vertical solid line will be accessible to
LEP2$^3$. The region on the left of
the vertical dashed line will be explorable at TeV33$^3$.
In the region delimited by the closed dashed line, the
neutralino relic abundance $\Omega_\chi h^2$ may exceed the value 0.1. \label{fig:mod1}}
\end{figure}

\begin{figure}[t]
\centerline{\psfig{figure=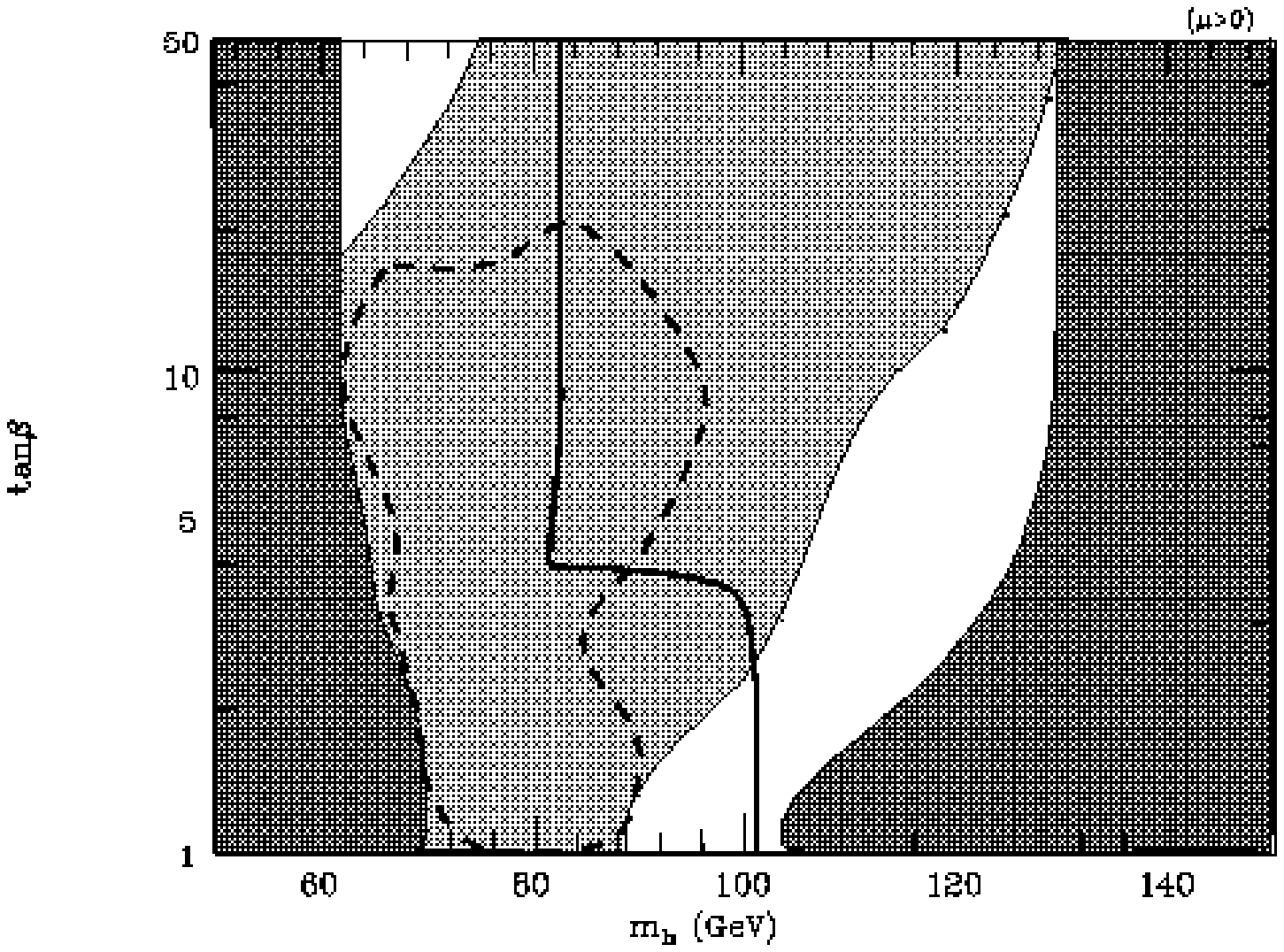,width=4.65in,bbllx=36bp,bblly=302bp,bburx=576bp,bbury=676bp,clip=}}
\fcaption{The same configurations of Fig. 4 are 
displayed in the $m_h$--$\tan\beta$ plane, within the gray area. 
The dark regions are excluded by current
LEP searches$^3$ or by theoretical arguments.
Configurations which provide $\Omega_\chi h^2 > 0.1$
fall within the region delimited by the closed dashed line.
The region on the left of the solid line 
will be accessible to LEP2$^3$. \label{fig:mod2}}
\end{figure}

\begin{figure}[t]
\centerline{\psfig{figure=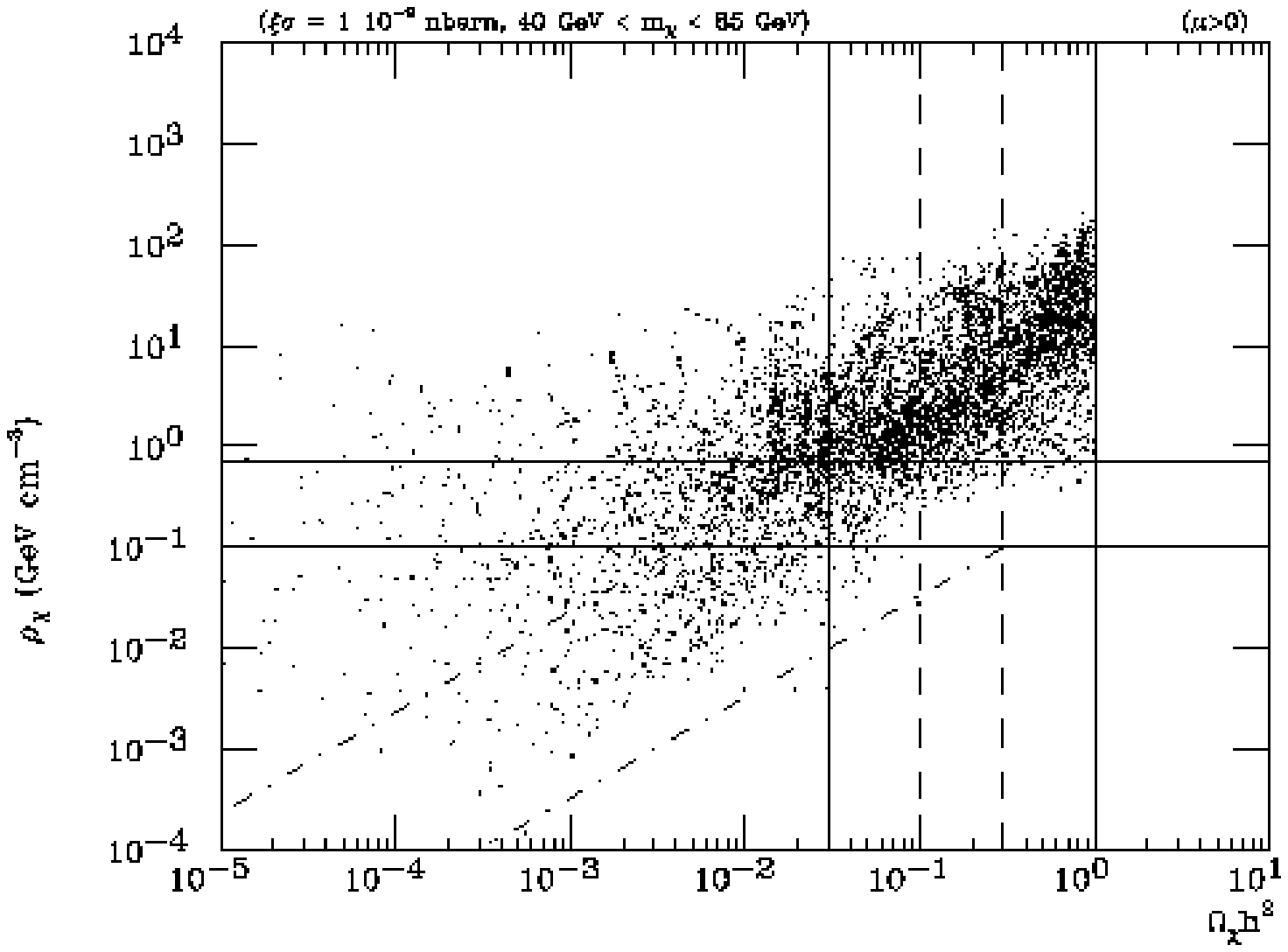,width=4.65in,bbllx=36bp,bblly=270bp,bburx=576bp,bbury=625bp,clip=}}
\fcaption{The neutralino local density $\rho_\chi$, calculated
for $[\xi \sigma^{(\rm nucleon)}_{\rm scalar}]_{\rm expt} =
1 \cdot 10^{-9}$ nbarn and $\rho_l = 0.5$ GeV cm$^{-3}$, is plotted versus the
neutralino relic abundance $\Omega_\chi h^2$.
For the value of $[\xi \sigma^{(\rm nucleon)}_{\rm scalar}]_{\rm expt}$
employed here, the neutralino mass is restricted to
the range 40 GeV $\lsim m_\chi \lsim 85$ GeV, as
obtained from the closed contour in Fig. 1.
The two horizontal lines denote the physical range  
for the local density of non--baryonic dark matter.
The two solid vertical lines denote the physical band for 
$\Omega h^2$, and the two dashed lines give the preferred band 
for the cold dark matter contribution to $\Omega$. The two
tilted dot--dashed lines denote the band where linear rescaling 
procedure for the local density is usually applied. \label{fig:mod3}}
\end{figure}

During the orbital motion of the Earth around the Sun, 
the direction of the velocity of the relic particles 
with respect to the detector changes as a function of time, and
this induces a time dependence in the differential detection
rate, i.e.\cite{freese,damamod}
$S(E,t) = S_0 (E) + S_m (E) \cos [\omega (t-t_0)]$,
where $\omega = 2\pi/365$ days and $t_0 = 153$ days
(June 2$^{\rm nd}$). $S_0 (E)$ is the average
(unmodulated) differential rate and $S_m (E)$ is the
modulation amplitude of the rate.
The relative importance of $S_m (E)$ with respect to $S_0 (E)$ for a given
detector, depends both on the mass of the dark matter particle and on the value of the
recoil energy where the effect is looked at. Typical values of $S_m (E)/S_0 (E)$
for a NaI detector are from a few percent up to roughly 15\%,
for neutralino masses of the order of 20--80 GeV and recoil energies
below 8--10 KeV. 

The DAMA/NaI Collaboration reported on an analysis of a 
collection of data over an exposure of 4549 Kg $\times$ days,
obtained with an experimental set-up
consisting of nine 9.70 Kg NaI(Tl) detectors\cite{damamod}. 
The extraction of a possible signal is obtained by employing 
a maximum likelihood method applied to a 
binning in the recoil energy of the daily counts per detector.  
When interpreted in terms of a modulation 
signal due to a WIMP of mass $m_{\chi}$ and elastic cross 
section $\sigma^{(\rm nucleon)}_{\rm scalar}$, 
the data of Ref.\cite{damamod} single 
out (at 90\% C.L.) the region of Fig. 1 which is delimited by a closed 
contour (hereafter defined as region $R_m$). 
The occurrence of region $R_m$ as a domain relevant for a possible  
modulation effect will necessarily require further investigation with 
much higher statistics. This point has clearly been stressed in 
Ref.\cite{damamod}. Meanwhile, it is very interesting to analyze the
possible implications that can be inferred from this preliminary result. 
Specifically, a number of questions deserve to be answered:
a) what would be the features of a 
neutralino to satisfy the prerequisites 
of region $R_m$; b) would any other experimental 
search for relic neutralinos be able to 
investigate the region $R_m$; c) are neutralino configurations of
region $R_m$ accessible to accelerator searches 
in the near future; d) what are the cosmological and astrophysical
implications of relic neutralinos whose scalar elastic cross section
is compatible with region $R_m$?

We start our analysis by comparing the result of Ref.\cite{damamod}
with our calculation within the MSSM scheme. This comparison is shown in
Fig.1, where we observe that many supersymmetric configurations
provide a value of $\xi \sigma^{\rm (nucleon)}_{\rm scalar}$ compatible
with the region $R_m$ (we will call hereafter set $S$ the supersymmetric
configurations compatible with region $R_m$).
As it was shown in Ref.\cite{pinning}, many
of these configurations provide sizeable muon fluxes from the Earth
and the Sun, reachable by the foreseeable improvements in neutrino telescopes.
This result is interesting since it suggests the possibility to have
an independent tool of exploration of the same configurations of set $S$.
In the case of the indirect signal from the Earth, it is already possible to 
exclude part of the configurations of set $S$ since they provide muon fluxes
exceeding the Baksan and Macro Collaborations limits\cite{pinning}. We therefore
include in our following considerations only configurations which are not in conflict
with indirect searches at neutrino telescopes (we define this subset as set $T$).

Let us now consider the features of the supersymmetric configurations 
of set $T$.  Fig.4 shows, within the gray area, the neutralino mass ranges compatible 
with set $T$, for different values of $\tan\beta$.
The dark area on the left side of the figure is
excluded by current LEP data\cite{LEP}. 
The region on the left of the
vertical solid line around $m_\chi \simeq 50$ GeV is the
region explorable by LEP2\cite{LEP}.
We notice that, as far as the neutralino sector
is concerned, LEP will be able to investigate only marginally
the region of parameter space singled out by set $T$.
A better chance to explore the neutralino mass range
compatible with set $T$ is given by the future upgrades
at the Tevatron and by LHC. As an example, the region which
extends up to the vertical dashed line is the mass region
which will be possibly explored by TeV33\cite{LEP}.

Fig.5 shows the properties of the configurations of set $T$ 
with respect to the lightest Higgs mass $m_h$, a parameter
which is crucial in establishing the size of both
direct and indirect detection signals. 
The gray area in Fig.5 corresponds to values of $m_h$ and $\tan\beta$
compatible with set $T$.
The dark regions are excluded by current LEP searches\cite{LEP}
(on the left of the plot) or by theoretical arguments (on the right side).
In Fig.5 it is also reported the region which will be accessible to LEP2\cite{LEP}. 
A large portion of the region
compatible with the modulation analysis will be covered by the
LEP2 searches. In particular, all the region for 
$\tan\beta \lsim 3$ will be analyzed. 

In Ref.\cite{pinning} it was shown that a large number of 
configurations belonging to set $T$ provide a large value
of the neutralino relic abundance. These configurations
are very appealing from the point of view of dark matter,
and they are shown in Fig.4,5 where the closed dashed line
contains the portion of the parameter space where 
$\Omega_\chi h^2$ may exceed the value 0.1.
In order to investigate the cosmological and astrophysical properties 
of the configurations of set $T$, we present an
analysis which is meant to obtain from the experimental data the relevant 
cosmological implications for relic neutralinos
in the most direct way. We adopt the following procedure\cite{extending}:
1) we evaluate $\sigma^{(\rm nucleon)}_{\rm scalar}$ and 
$\Omega_\chi h^2$ by varying the supersymmetric 
parameters;
2) for any value of 
$[\rho_\chi \sigma^{(\rm nucleon)}_{\rm scalar}]_{\rm expt}=\rho_l [\xi
\sigma^{(\rm nucleon)}_{\rm scalar}]_{\rm expt}$ 
compatible with 
the experimental region $R_m$ we calculate
$\rho_\chi = [\rho_\chi \sigma^{(\rm nucleon)}_{\rm scalar}]_{\rm expt} / 
\sigma^{(\rm nucleon)}_{\rm scalar}$
and restrict the values of $m_\chi$ to stay inside the region $R_m$.
3) The results are displayed in a scatter plot in the plane 
$\rho_\chi$ vs. $\Omega_\chi h^2$.

One example of our results is given in Fig.6 for
$[\xi \sigma^{(\rm nucleon)}_{\rm scalar}]_{\rm expt} = 1 \cdot 10^{-9}$ nbarn
($\rho_l=0.5$  GeV cm$^{-3}$ is used).
The two horizontal lines denote the physical range  
0.1 GeV cm$^{-3}$  $< \rho_l <$ 0.7 GeV cm$^{-3}$ for the local density 
of non--baryonic dark matter\cite{extending}.
The solid vertical lines  denote the physical band for $\Omega h^2$:
$0.03 \lsim \Omega h^2 \lsim 1$, and the two 
dashed lines give the favored band for the cold dark matter 
contribution to $\Omega$: 
$0.1 < (\Omega h^2)_{\rm CDM} < 0.3$\cite{extending}. The two tilted 
dot--dashed lines denote the band where linear rescaling procedure for the 
local density is usually applied. 
     With the aid of this kind of plot we can classify the supersymmetric 
configurations belonging to region $R_m$ into various categories.
Configurations whose representative points fall above the maximum 
value $\rho_\chi = 0.7$ GeV cm$^{-3}$ 
have to be excluded (those providing 
an $\Omega_\chi h^2 > 1$ are already disregarded in the plot).
Among the allowed configurations, those falling 
in the region inside 
both the  horizontal and solid vertical lines are very 
appealing, since they would represent situations where the neutralino 
could have the role of a dominant cold dark matter component; even more so,
if the representative points fall 
in the subregion inside the vertical 
band delimited by dashed lines. Configurations which fall inside 
the band delimited by the tilted dot--dashed lines denote situations 
where the neutralino can only provide a fraction of the cold dark 
matter both at the level of local density and at the level of the 
average $\Omega$. Configurations above the upper dot--dashed line and below 
the upper solid horizontal line would imply an unlikely special 
clustering of neutralinos in our halo as compared to their average 
distribution in the Universe.

\vspace{-7pt}
\section{Conclusions}
\vspace{-10pt}
In this paper we have reported the most recent calculations of the
direct and indirect detection rates of relic neutralinos in the
framework of the Minimal Supersymmetric Standard Model. We have
shown that the theoretical estimates of the detection rates may be
at the level of the present experimental sensitivities of low--background
detectors and neutrino telescopes. For many supersymmetric configurations,
and for median values of the astrophysical parameters which enter in the
calculations of the detection rates, the predicted signals may already
exceed the present experimental bounds. This shows that the different
experimental efforts to search for relic particles are potentially able 
to deeply investigate the possibility that neutralino is a component
of the dark matter of the Universe. An interesting preliminary analysis of the
DAMA/NaI Collaboration has shown that their data are compatible, at 90\% C.L.,
with a modulation signal of the direct detection rate.
The features of a neutralino able to satisfy the prerequisites of this signal 
have been analyzed and it has been
shown that many configurations are compatible with a dark matter scenario
where the neutralino is the major component, both on galactic and cosmological
scales. However, we have to remind here that the occurrence of a possible
modulation effect will necessarily require further investigations with
much higher statistics. This project is currently under way.

\end{document}